**Title:** Computer Vision for Increased Operative Efficiency via Identification of Instruments in the Neurosurgical Operating Room: A Proof-of-Concept Study

**Authors:** Tanner J. Zachem[1,2] BSE, Sully Chen BS[1], Vishal Venkatraman MHSc[1], David AW Sykes BA[1], Ravi Prakash MS[2], Koumani W. Ntowe BS[1], Mikhail A. Bethell MS[1], Samantha Spellicy MD PhD[1], Alexander D. Suarez MD[1], Weston Ross PhD[1], Patrick J. Codd MD[1,2]

[1]Department of Neurosurgery, Duke University School of Medicine, Durham, NC, USA
[2]Department of Mechanical Engineering and Materials Science, Duke University, Durham, NC, USA

**Corresponding Author:**
Patrick J. Codd, MD
Department of Neurosurgery, Duke University School of Medicine
Email: patrick.codd@duke.edu

**Disclosures:** None related to the described works.

**Funding:** This work was partially supported by the Division of Graduate Education of the National Science Foundation under the NSF Traineeship in the Advancement of Surgical Technologies, Award #2125528.

**Data Availability:** Data is openly available through The Open Science Framework: https://doi.org/10.17605/OSF.IO/BCQK2.

## Abstract

*Objectives*

Computer vision (CV) is a field of artificial intelligence that enables machines to interpret and understand images and videos. CV has the potential to be of assistance in the operating room (OR) to track surgical instruments. We built a CV algorithm for identifying surgical instruments in the neurosurgical operating room as a potential solution for surgical instrument tracking and management to decrease surgical waste and opening of unnecessary tools.

*Methods*

We collected 1660 images of 27 commonly used neurosurgical instruments. Images were labeled using the VGG Image Annotator and split into 80% training and 20% testing sets in order to train a U-Net Convolutional Neural Network using 5-fold cross validation.

*Results*

Our U-Net achieved a tool identification accuracy of 80-100% when distinguishing 25 classes of instruments, with 19/25 classes having accuracy over 90%. The model performance was not adequate for sub classifying Adson, Gerald, and Debakey forceps, which had accuracies of 60-80%.

*Conclusions*

We demonstrated the viability of using machine learning to accurately identify surgical instruments. Instrument identification could help optimize surgical tray packing, decrease tool usage and waste, decrease incidence of instrument misplacement events, and assist in timing of routine instrument maintenance. More training data will be needed to increase accuracy across all surgical instruments that would appear in a neurosurgical operating room. Such technology has the potential to be used as a method to be used for proving what tools are truly needed in each type of operation allowing surgeons across the world to do more with less.

## Introduction

Technological automations and advancements have increasingly enhanced the efficiency and safety of surgical experience. Innovations such as augmented reality, robotics, and artificial intelligence have supplemented traditional methods in surgery.[1-3] These technologies have allowed surgeons to better plan trajectories for surgical approaches, increased the accuracy of techniques such as pedicle screw placement, predicted the location of epileptic foci in the brain with better accuracy than humans, and helped identify patients who are at high risk for postoperative complications.[3] These technologies are leading to improved patient outcomes, decreased operating times, and enhanced training for surgical residents.[4,5]

However, the addition of equipment and new tools in the OR has led to clutter, posing a hazard and inability to keep track of supplies in the middle of a surgery, while also increasing the number of items people feel are necessary for an operation.[6,7] Instrument counting and tracking in the operating room remains a mostly manual task when it is employed, and subject to mistakes from human error. There exists an opportunity for better instrument tracking automatization and optimization in the operating room, as well as quantifying which tools are most utilized and necessary for a procedure.

Computer vision (CV) is a field of artificial intelligence and computer science that enables machines to interpret and understand images and videos and has great potential to be of assistance in the OR. Prior research on CV in the OR has demonstrated its utility in tracking surgeon movements, allowing for assessment of surgeon efficiency,[8,9] evaluation of surgeon performance and safety,[10] and assessing specific movements performed during a procedure.[11] CV has also allowed for identification of polyps and cancers in general surgery.[12,13]

Using computer vision in surgery is also practical given the ubiquity of cameras in the OR, such as those found in endoscopes, laparoscopes, and cameras placed on overhead arms to record surgical video.[13] Therefore, we sought to build a CV model to accurately identify and track instruments in the operating room. We therefore collected a set of commonly used surgical instruments from the neurosurgical toolsets at our academic medical center, created a small pixel-wise segmentation dataset of these instruments, and trained a deep-learning based computer vision model.

## Methods

*Instrument, Data Collection, and Data Preparation*

We obtained a set of 27 common neurosurgical instruments and supplies used by the lead surgeon in this study (PJC). The unique instruments analyzed included various forceps (Adson, bayonet, DeBakey, Gerald), clamps (allis, sponge stick, right angle, tonsil, mosquito), bone curettes and rongeurs (Leksell, Kerrison, small bone curette, periosteal elevator), retractors (Cobb, Weitlaner, Army-Navy), disposables (syringes, irrigation bulbs, neuro pattie sponges, raytec sponges), and other instruments (bovie cautery, needle driver, scalpel, Metzenbaum scissors, suction tips, clip appliers, and Raney clip appliers).

Multiple images were obtained for each instrument, taken from multiple angles and multiple orientations, using the wide lens (f/1.5 focal length, 12-megapixel resolution) camera of an iPhone 13 Pro (Apple Inc., Cupertino, CA). A minimum of 30 images were taken for each item, and instruments with articulating joints or other manipulations had images taken of each geometry. Each instrument was photographed while placed by itself on green surgical towels to mimic the surface of a typical OR supply table. Standard overhead lighting was used to mimic the lighting conditions of the OR. Images were imported into the VGG Image Annotator,[14] which was used to draw bounding boxes around the instrument in each photo as seen in **Figure 1**.

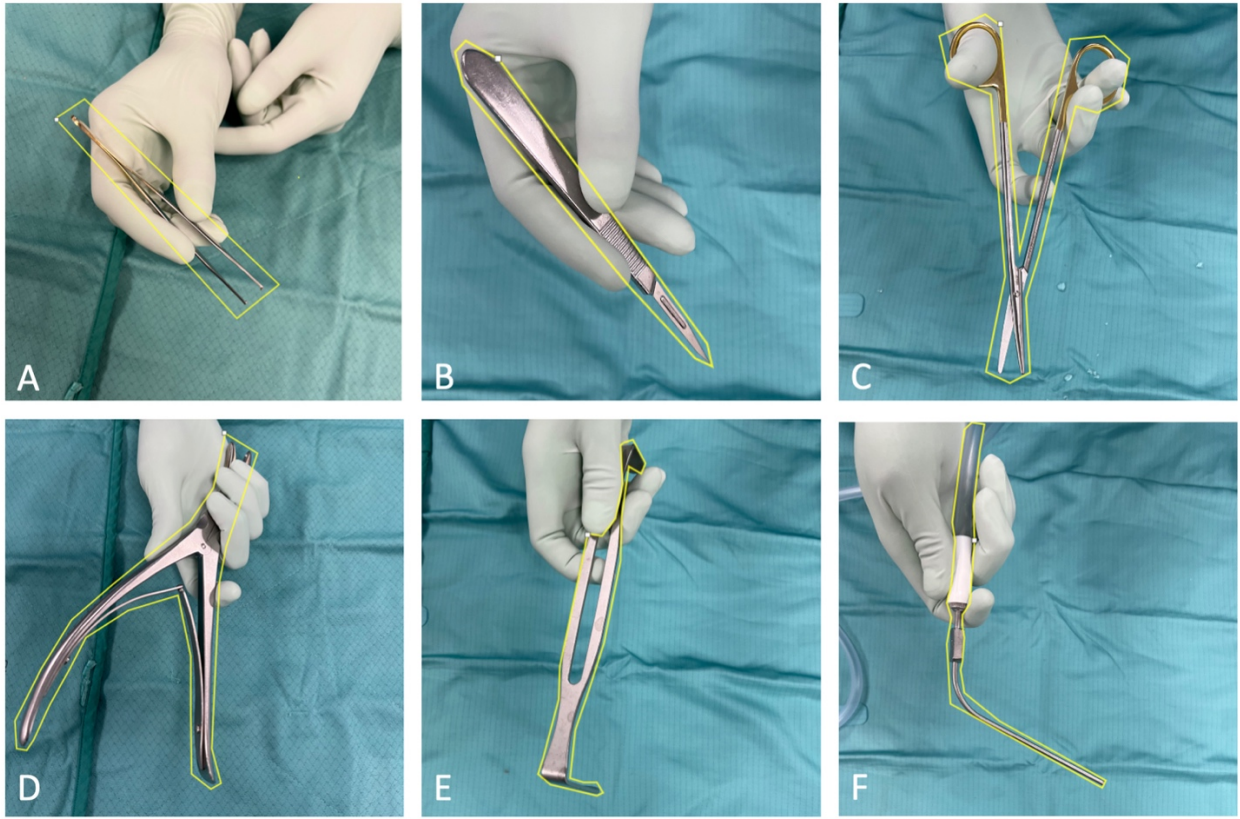

*Figure 1: Bounding boxes drawn on images. A) Adson Forceps B) Scalpel C) Metzenbaum Scissors D) Leksell Rongeur E) Army-Navy Retractor F) Suction*

The images were then labeled with their ground truth (actual instrument) by the study team. For purposes of the computer vision model, we automatically labeled any portion of the image not within the bounding box as "not a tool."

*Architecture, Training, and Testing*

Because our goal was to perform pixel-level classification of surgical instruments, we opted to use a U-Net,[15] a state-of-the-art convolutional neural network model for medical image segmentation. To implement the U-Net, we cloned an open-source implementation of the architecture and adjusted parameters as needed for our task. To see details on our optimization, training, and testing parameters, see **Supplemental Materials**. We split our dataset into five folds for cross validation. In this manner, the model tunes its parameters using 80% of the images and then is tested on the remaining 20% images without having access to the ground truth of what the image is classified as.

We evaluated two metrics: accuracy of classification and intersection-over-union (IoU). Accuracy measures the percentage of correct classifications over the total number of test images. Whereas IoU compares the number of pixels contained in the predicted region to the pixels that make up the ground truth of the surgical instrument. The ground truth is represented by the manually segmented tool. IoU is calculated by dividing the area of overlap between the prediction and ground truth by the sum of the two areas and ranges from 0.0 to 1.0, where 1.0 represents best performance (complete overlap between prediction and ground truth).

Intersection-over-union is computed by comparing the set of pixels that form the predicted region (cardinality) of the surgical instrument to the ground truth set of pixels that constitute the surgical instrument. The IoU score is generated by dividing the cardinality of the intersection of these two sets by the cardinality of the union of these two sets, yielding a number between 0.0 and 1.0, with 1.0 representing the best possible performance.

## Results

*Study Dataset and Data Pre-Processing*

A total of 1660 images were taken to build the dataset for this study. The number of images of each instrument is listed in **Table 1** (n=1328) and then is tested on the remaining 20% (n=332) images without having access to the ground truth. Basic preprocessing steps for a U-Net were applied, such as dimension reduction, image normalization, cropping, and basic scaling for machine learning.

Table 1: Model Performance by Instrument

| Instrument | Sample Size | Accuracy (mean ± SD) | IoU (mean ± SD) |
|---|---|---|---|
| Adson Forceps | 44 | 63.64% ± 2.52% | 0.7135 ± 0.2326 |
| Allis Clamp | 81 | 86.42% ± 4.71% | 0.6504 ± 0.2722 |
| Army Navy Retractor | 42 | 88.10% ± 5.60% | 0.6264 ± 0.2239 |
| Bayonet Forceps | 61 | 91.80% ± 8.83% | 0.6847 ± 0.1991 |
| Bone Curette | 60 | 85.00% ± 6.24% | 0.5939 ± 0.2453 |
| Bovie Cautery | 80 | 93.75% ± 10.54% | 0.5591 ± 0.2674 |
| Clip Applier | 70 | 94.29% ± 7.75% | 0.7332 ± 0.1476 |
| Cobb Retractor | 47 | 87.23% ± 7.06% | 0.6422 ± 0.2235 |
| Debakey Forceps | 25 | 68.00% ± 26.51% | 0.5385 ± 0.3348 |
| Gerald Forceps | 35 | 74.29% ± 15.03% | 0.5790 ± 0.2675 |
| Irrigation Bulb | 36 | 100.00% ± 0.00% | 0.8566 ± 0.0776 |
| Kerrison Rongeur | 99 | 90.91% ± 4.09% | 0.6342 ± 0.1826 |
| Leksell Rongeurs | 43 | 93.02% ± 12.67% | 0.7136 ± 0.1867 |
| Metzenbaum Scissors | 62 | 98.39% ± 3.29% | 0.7054 ± 0.0861 |
| Mosquito Clamp | 26 | 92.31% ± 9.73% | 0.7327 ± 0.1655 |
| Needle Driver | 59 | 86.44% ± 9.13% | 0.5996 ± 0.2193 |
| Neuro Pattie Sponges | 24 | 91.67% ± 14.91% | 0.6659 ± 0.3120 |
| Periosteal Elevator | 93 | 96.77% ± 2.40% | 0.4381 ± 0.3307 |
| Raney Clip Applier | 43 | 93.02% ± 5.97% | 0.7579 ± 0.1947 |
| Raytec Sponge | 11 | 90.91% ± 12.03% | 0.8029 ± 0.1893 |
| Right Angle Forceps | 72 | 95.83% ± 5.33% | 0.6729 ± 0.1354 |
| Scalpel | 99 | 92.93% ± 4.55% | 0.7414 ± 0.1691 |
| Sponge Stick | 64 | 95.31% ± 6.92% | 0.6948 ± 0.1600 |
| Suction | 91 | 95.60% ± 4.73% | 0.5580 ± 0.1927 |
| Syringe | 141 | 97.87% ± 2.38% | 0.6423 ± 0.3121 |
| Tonsil Forceps | 83 | 97.59% ± 3.08% | 0.4263 ± 0.3181 |
| Weitlaner Retractor | 69 | 92.75% ± 6.54% | 0.6733 ± 0.1667 |

*Predicting Instrument Type Using U-Net*

The U-Net model was trained over 5 different 80%:20% train/test splits to identify surgical instruments using a convolutional neural network algorithm. This was done with 5-fold cross validation to determine if overfitting is present in the model. The average performance by accuracy & IOU for each of these five tests is shown in **Table 1**. The accuracy in correctly identifying instruments in the test set ranged from 63.64% for Adson Forceps to 100% for Irrigation Bulbs. Nineteen of the instruments were identified with an average accuracy of at least 90%, while five instruments were identified with an accuracy between 80% to 90%. Three of the types of forceps (Adson, Gerald, & Debakey) had average identification accuracies below 80%. The IOU for pixel-level accuracy ranged from an average of 0.4263 for Tonsil Forceps to 0.8566 for Irrigation Bulb.

Examples of heat maps generated by our model are shown in **Figure 2**.

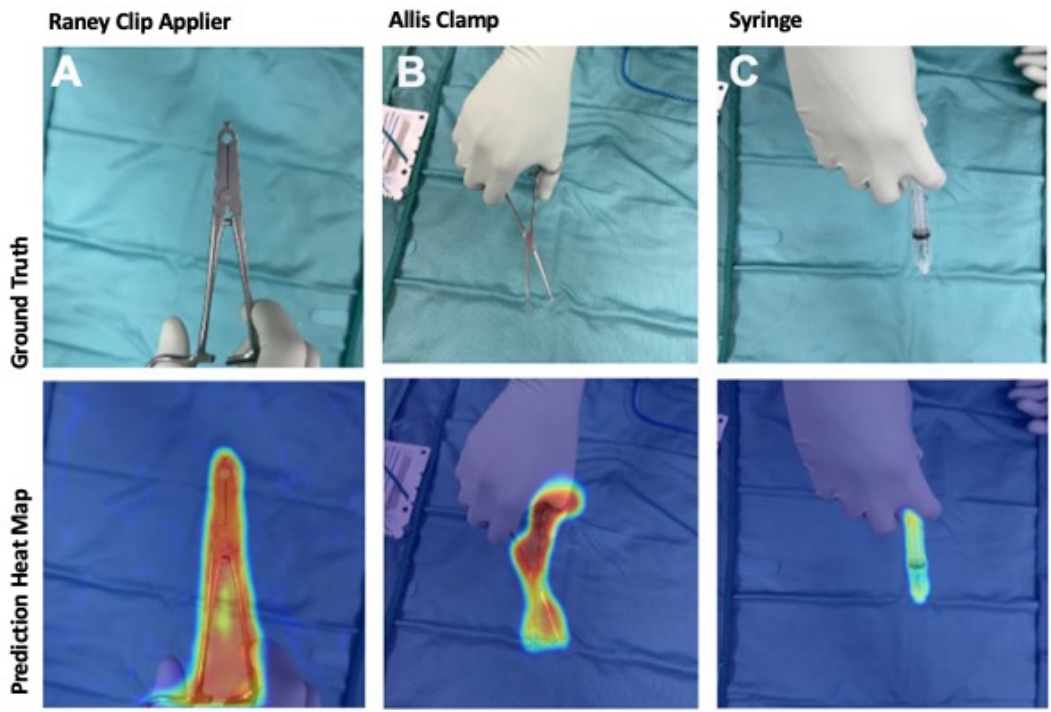

*Figure 2: - Prediction efficacy of instrument identification and recognition. Heatmaps demonstrating the correctly identified instrument area and name for A) Raney Clip Applier B) Allis Clamp C) Syringe*

These heat maps demonstrate the model's ability to select which pixels in an image are likely part of the instrument to be identified. The heat map demonstrates which pixels in the image the model identified to be a part of a given tool, and the color of the heat map ranges from blue (cool) to red (hot) and demonstrates the model's confidence. The darker the color (higher temperature) implies a higher level of confidence.

## Discussion

We present here a novel proof-of-concept study of a CV algorithm used to identify and classify surgical instruments, that has potential to quantitatively describe which tools are used most and needed for certain operations and which tools are extra and not needed. Our model consisting of a series of convolutional neural networks in the U-Net architecture was able to accurately identify instruments presented in isolation, with most tools identified at 80-100% accuracy. These results demonstrate the potential of CV algorithms as a method of instrument tracking in surgery. The model presented here represents the first step of a larger effort to integrate CV into the operating room as an additional instrument to allow for increased efficiency, save hospital resources, and prevent mistakes. Therefore, this model can help quantify what can be done with certain tools and allow hospitals to improve their economic and ecological footprints.

Convolutional neural networks with a U-Net framework have been used in prior studies to solve a variety of biomedical imaging tasks, including identifying lesions in the retina,[16] segmenting tumors from brain MRIs,[17] and counting individual cells from microscopy images.[18] While U-Net based models in these studies have shown good accuracy, authors have noted key limitations in this software, mainly that they identify objects with approximate outlines and can have difficulty in differentiating smaller areas of detail.[17,19] While this limitation was generally not an issue with our dataset, as surgical instruments tend to look quite different from one another, this issue could be seen in our models' ability to identify distinct types of forceps. Instruments such as the Adson, Debakey, and Gerald forceps are all straight forceps that are differentiated by the type of teeth or tissue grip at the tip of the instrument, which could explain why our model had lower accuracy in identifying these. Overall, however, our U-Net algorithm showed high overall accuracy for most instruments and could be used to solve the problems we identified in the modern operating room.

A key issue in an operating room is clutter and lack of efficiency in instrument use.[6,7] For a given operation, a tray with dozens of instruments is opened for potential use, but only a handful of instruments are picked out and used by the surgical team. However, the entire tray must be reprocessed, adding to costs and in smaller care centers, decreasing the number of operations possible per day. One center conducted a study of surgical tray optimization and found that eliminating unused instruments from trays saved up to $500,000 a year due to saved costs from sterile processing, packing, and counting time, and replacing lost or damaged instruments.[20,21] Additionally, with instrument loss or misplacement occurring at a rate of about 1%,[21] the loss of surgical instruments has been estimated to cost up to $350,000 for a hospital.[22] Our model looks to quantify which tools are used most often and truly needed for the cases. Therefore, a hospital with a more limited supply of tools, could utilize the model to optimize tool tray deployment.

Retained foreign objects are a critical error that can occur after surgery, and describes instruments, sponges, or other items being left inside patients.[23,24] While rare, these events can cause great morbidity or even mortality to patients and require reoperation to remove the object. These errors are often the result of human error, such as miscounts at the end of an operation or lack of visual identification.

Another key application for surgical tool tracking is in measuring surgeon skill and progression. Preliminary studies to track hand motions and instrument position during surgery have shown promise in identifying differences in technique between different surgeons.[8,9,25] If refined, such technology has the potential to measure progress in surgical skill for trainees, especially in fields such as cerebrovascular or skull base neurosurgery, which require precise and fine movements to avoid damage to critical structures.

We envision that a CV algorithm tracking instrument used in real time could manage all these surgical problems. An algorithm to analyze videos over multiple surgeries could allow surgeons to identify which instruments they most often use, allowing for individualized tray optimization. CV could also ensure that instruments are returned to a tray at the end of a case and help prevent the loss of expensive surgical instruments. While surgical counts are performed by OR staff to ensure that every object used during a procedure has left the patient and been returned to the tray or set aside for disposal, an algorithm that counts objects could serve as an additional check to prevent the critical error of retained foreign objects. Furthermore, increased automation could lead to decreased total OR and turnover time and therefore an increase in surgical cases in the same operating room. These increases in case capacity per room could result in a substantial increase in a hospital's overall case volume with the same resources.

Furthermore, in hospitals with limited resources, this technology poses to prove what resources are truly needed for each operation, allowing for more efficient use of fewer tools, and preventing significant waste and/or re-sterilization that would delay other operations.

Our model has some key limitations that we plan to address in future iterations. The primary limitation lies in the setting in which images were obtained. A typical OR has bright spot lighting and dozens of instruments and instruments laid out on a table as well as multiple people moving around during surgery, while this study used images of single instruments placed in a non-clinical setting. Our tools were also cleaned prior to obtaining images, so our model does not account for the real-world scenario in which tools are often covered in blood or inserted deep into body cavities. Also, our dataset is limited and does not include the entire scope of instruments that could be used during surgery.

Future directions for this research include obtaining data directly from an OR during surgery, both by taking pictures of the operating table as instruments are being organized, as well as live video of instruments being moved and manipulated by surgeons and their assistants. Training an algorithm based on this dataset would significantly increase the usability of CV in a real-world setting. Future iterations of our model could also include optimization work to allow for the differentiation of similar looking instruments, such as the forceps that our model had trouble identifying.

**Conclusion**

We created a U–Net based computer-vision algorithm and demonstrated its viability to accurately identify instruments commonly used in neurosurgery. For all but 3 of the 27 types of instruments we tested, we achieved greater than 80% accuracy. More training data will be needed to increase accuracy across all surgical instruments that would appear in a neurosurgical operating room. Such technology and its data stream have the potential to be used as a method to track surgical instruments, optimize data around instrument usage and instrument supply in the operating room, evaluate surgeon performance, help with instrument inventory and organization, prevent incidents such as retained foreign objects, and quantitatively describe how to do more with less.

# References

1. Verhey JT, Haglin JM, Verhey EM, Hartigan DE. Virtual, augmented, and mixed reality applications in orthopedic surgery. *Int J Med Robot.* 2020;16(2):e2067.
2. Andras I, Mazzone E, van Leeuwen FWB, et al. Artificial intelligence and robotics: a combination that is changing the operating room. *World J Urol.* 2020;38(10):2359-2366.
3. Mofatteh M. Neurosurgery and artificial intelligence. *AIMS Neurosci.* 2021;8(4):477-495.
4. Marcus HJ, Hughes-Hallett A, Kwasnicki RM, Darzi A, Yang GZ, Nandi D. Technological innovation in neurosurgery: a quantitative study. *J Neurosurg.* 2015;123(1):174-181.
5. Ho A, Khan YR, Whitney E, Alastra AJ, Siddiqi J. The Influence of Intraoperative Technology on Neurosurgery Training. *Cureus.* 2019;11(9):e5769.
6. Alarcon A, Berguer R. A comparison of operating room crowding between open and laparoscopic operations. *Surg Endosc.* 1996;10(9):916-919.
7. Ofek E, Pizov R, Bitterman N. From a radial operating theatre to a self-contained operating table. *Anaesthesia.* 2006;61(6):548-552.
8. Ganni S, Botden S, Chmarra M, Goossens RHM, Jakimowicz JJ. A software-based tool for video motion tracking in the surgical skills assessment landscape. *Surg Endosc.* 2018;32(6):2994-2999.
9. Zhang M, Cheng X, Copeland D, et al. Using Computer Vision to Automate Hand Detection and Tracking of Surgeon Movements in Videos of Open Surgery. *AMIA Annu Symp Proc.* 2020;2020:1373-1382.
10. Kennedy-Metz LR, Mascagni P, Torralba A, et al. Computer Vision in the Operating Room: Opportunities and Caveats. *IEEE Trans Med Robot Bionics.* 2021;3(1):2-10.
11. Luongo F, Hakim R, Nguyen JH, Anandkumar A, Hung AJ. Deep learning-based computer vision to recognize and classify suturing gestures in robot-assisted surgery. *Surgery.* 2021;169(5):1240-1244.
12. Kitaguchi D, Takeshita N, Hasegawa H, Ito M. Artificial intelligence-based computer vision in surgery: Recent advances and future perspectives. *Ann Gastroenterol Surg.* 2022;6(1):29-36.
13. Chadebecq F, Vasconcelos F, Mazomenos E, Stoyanov D. Computer Vision in the Surgical Operating Room. *Visc Med.* 2020;36(6):456-462.
14. Dutta A, Zisserman A. The VIA Annotation Software for Images, Audio and Video. Proceedings of the 27th ACM International Conference on Multimedia; 2019; Nice, France.
15. Yin XX, Sun L, Fu Y, Lu R, Zhang Y. U-Net-Based Medical Image Segmentation. *J Healthc Eng.* 2022;2022:4189781.
16. Kundu S, Karale V, Ghorai G, Sarkar G, Ghosh S, Dhara AK. Nested U-Net for Segmentation of Red Lesions in Retinal Fundus Images and Sub-image Classification for Removal of False Positives. *J Digit Imaging.* 2022;35(5):1111-1119.
17. Zhao L, Ma J, Shao Y, Jia C, Zhao J, Yuan H. MM-UNet: A multimodality brain tumor segmentation network in MRI images. *Front Oncol.* 2022;12:950706.
18. Falk T, Mai D, Bensch R, et al. U-Net: deep learning for cell counting, detection, and morphometry. *Nat Methods.* 2019;16(1):67-70.
19. Wan C, Wu J, Li H, et al. Optimized-Unet: Novel Algorithm for Parapapillary Atrophy Segmentation. *Front Neurosci.* 2021;15:758887.

20. Farrelly JS, Clemons C, Witkins S, et al. Surgical tray optimization as a simple means to decrease perioperative costs. *J Surg Res.* 2017;220:320-326.
21. Zhu X, Yuan L, Li T, Cheng P. Errors in packaging surgical instruments based on a surgical instrument tracking system: an observational study. *BMC Health Serv Res.* 2019;19(1):176.
22. Rodrigues M, Mayo M, Patros P. OctopusNet: Machine learning for intelligent management of surgical tools. *Smart Health.* 2022;23:100244.
23. Steelman VM, Shaw C, Shine L, Hardy-Fairbanks AJ. Unintentionally Retained Foreign Objects: A Descriptive Study of 308 Sentinel Events and Contributing Factors. *Jt Comm J Qual Patient Saf.* 2019;45(4):249-258.
24. Lincourt AE, Harrell A, Cristiano J, Sechrist C, Kercher K, Heniford BT. Retained foreign bodies after surgery. *J Surg Res.* 2007;138(2):170-174.
25. Ahmidi N, Hager GD, Ishii L, Fichtinger G, Gallia GL, Ishii M. Surgical task and skill classification from eye tracking and tool motion in minimally invasive surgery. *Med Image Comput Comput Assist Interv.* 2010;13(Pt 3):295-302.

**Supplemental Materials**

To implement U-Net, we cloned an open-source implementation of the architecture and adjusted the classification head as needed for our task. We append a 28-channel output layer with a channel-wise SoftMax activation to compute the probability of each class at a per-pixel level. Due to the small amount of training data, we opt to fine-tune a U-Net encoder pretrained on ImageNet with a randomly initialized decoder. We train this architecture via the Adam optimizer ($\beta_1=0.9$, $\beta_2 = 0.999$, $\varepsilon=10^{-8}$) at a learning rate of 0.001 for 15000 iterations with a batch size of 128. We use mean-squared error loss with one-hot classification labels. Additionally, we apply a linear learning rate decay over the duration of the training, decaying from 0.001 to 0.0 on the last iteration. The choice of hyperparameters was taken from the default Adam initialization in PyTorch. Linear learning rate decay was chosen for its simplicity and previous empirical success. We apply random crops and resizes to the training dataset to augment the amount of training data.

*Testing*

Because our model is a pixel-level classification model, we classify individual tools by computing the predicted area occupied by each object in the image via a threshold; in other words, we mark any pixel with class-probability greater than some threshold as a pixel representing 1 unit area of that class, then we compute the area of each class in the image and assign the whole image to be the class with the maximum area.

Intersection-over-union is computed in a similar way to accuracy. First, the positive-prediction pixels are identified via a threshold value. These positive pixels constitute the set of pixels that form the predicted region of the surgical tool. We compare this set to the ground truth set of pixels that constitute the surgical tool. We generate the IoU score by dividing the cardinality (the number of pixels in the set), the number of pixels in the set, of the intersection of these two sets by the cardinality of the union of these two sets, yielding a number bound strictly between 0.0 and 1.0. This number yields an intuitive measure of how well the model classifies the tool on a pixel-level.

    To obtain a robust measure of model performance, we run five-fold cross-validation over the whole dataset. In other words, we create five 80:20 train-test splits (where each test split is mutually exclusive), train a model on each train set, and measure performance on each corresponding test set.